\documentclass[10pt,preprint]{aastex}

\begin{document}

\shortauthors{Luhman, Peterson, \& Megeath}
\shorttitle{Chamaeleon Brown Dwarf}

\title{Spectroscopic Confirmation of the Least Massive Known Brown Dwarf in Chamaeleon\altaffilmark{1}}

\author{K. L. Luhman\altaffilmark{2}, Dawn E. Peterson\altaffilmark{3,4}, 
\& S. T. Megeath\altaffilmark{2,3}}

\altaffiltext{1}
{Based on observations obtained at the Gemini Observatory, which is operated 
by the Association of Universities for Research in Astronomy, Inc., under 
a cooperative agreement with the NSF on behalf of the Gemini partnership: 
the National Science Foundation (United States), the Particle Physics and 
Astronomy Research Council (United Kingdom), the National Research Council
(Canada), CONICYT (Chile), the Australian Research Council (Australia), 
CNPq (Brazil) and CONICET (Argentina). 
This publication makes use of data products from the Two Micron All
Sky Survey, which is a joint project of the University of Massachusetts
and the Infrared Processing and Analysis Center/California Institute
of Technology, funded by the National Aeronautics and Space
Administration and the National Science Foundation.}

\altaffiltext{2}{Harvard-Smithsonian Center for Astrophysics, 60 Garden St.,
Cambridge, MA 02138, USA; kluhman, tmegeath@cfa.harvard.edu.}

\altaffiltext{3}
{Visiting Astronomer at the Infrared Telescope Facility, which is operated 
by the University of Hawaii under Cooperative Agreement no. NCC 5-538 with 
the National Aeronautics and Space Administration, Office of Space Science, 
Planetary Astronomy Program.}

\altaffiltext{4}{The University of Rochester, Department of Physics and 
Astronomy, Rochester, NY 14627, USA; dawnp@astro.pas.rochester.edu.}

\begin{abstract}

We present spectroscopy of two candidate substellar members of 
the Chamaeleon~I star-forming region.
The candidates, which were identified photometrically by Oasa, Tamura, \& 
Sugitani, have been observed at 1-2.5~\micron\ during commissioning of the 
Gemini Near-Infrared Spectrograph.
The late-type nature of one of the candidates, OTS~44, is confirmed through the 
detection of strong steam absorption bands. The other object, OTS~7,
exhibits no late-type features and is likely a background star or galaxy.
The gravity-sensitive shape of the $H$- and $K$-band continua demonstrate
that OTS~44 is a young, pre-main-sequence object rather than a field dwarf.  
We measure a spectral type of M9.5 for OTS~44 based on a comparison of its
spectrum to data for optically-classified young late-type objects. 
Because OTS~44 is the coolest and faintest object with confirmed membership
in Chamaeleon I, it is very likely the least massive known member of the
cluster. By comparing the position of OTS~44 on the H-R diagram to 
the evolutionary models of Chabrier \& Baraffe, we infer a mass of
$\sim0.015$~$M_\odot$. Although this estimate is uncertain by at least
a factor of two, OTS~44 is nevertheless one of the least massive 
free-floating brown dwarfs confirmed spectroscopically to date.

\end{abstract}

\keywords{infrared: stars --- stars: evolution --- stars: formation --- stars:
low-mass, brown dwarfs --- stars: pre-main sequence}

\section{Introduction}
\label{sec:intro}

Because brown dwarfs are relatively bright soon after birth and fade
precipitously thereafter, star-forming clusters are the most 
promising sites in which to search for the least massive brown dwarfs.
At a distance of 160-170~pc \citep{whi97,wic98,ber99}, the Chamaeleon~I 
cloud complex is one of the nearest major star formation regions
and therefore is a particularly attractive hunting ground.
A variety of methods have been used to identify candidate low-mass stars
and brown dwarfs in Chamaeleon~I, including objective prism spectroscopy at 
H$\alpha$ \citep{com99,com00,com04,nc99}, photometric variability \citep{car02},
X-ray emission \citep{com00,fl04}, and optical and infrared (IR) photometry
\citep{cam98,ots99,per00,per01,gk01,lm04,cc04}.
All spectroscopic observations of candidates from these surveys have been 
compiled and evaluated by \citet{luh04a}, with the exception of the
recent spectroscopy provided by \citet{com04}.
To date, the coolest confirmed members of Chamaeleon~I have spectral types 
of $\sim$M8, corresponding to masses of $\sim0.03$~$M_\odot$.

From the previous surveys of Chamaeleon, there remain several promising 
candidate brown dwarfs that lack spectroscopy and that should have masses 
as low as 0.01~$M_\odot$ if they are indeed late-type members of the
star-forming region.  
During the recent commissioning of the Gemini Near-Infrared Spectrograph 
(GNIRS), we obtained near-IR spectroscopy of the two faintest objects
from the compilation of candidates in \citet{luh04a},
sources 7 and 44 from \citet{ots99} (hereafter OTS~7 and 44). 
In this letter, we describe these observations (\S~\ref{sec:obs}), 
assess the membership of the candidates in Chamaeleon~I and measure 
the spectral type for the one confirmed member
(\S~\ref{sec:mem}), estimate the extinction, effective temperature, and 
bolometric luminosity of that source (\S~\ref{sec:ext}), 
and infer its mass from theoretical evolutionary models (\S~\ref{sec:mass}).

\section{Observations}
\label{sec:obs}

Near-IR spectroscopy was performed on the brown dwarf candidates OTS~44 and 7
during queue observations with GNIRS at Gemini South Observatory on the
nights of 2004 March 9 and 10, respectively. To facilitate the spectral 
classification of these objects, we also observed an optically-classified
late-type member of Chamaeleon~I, source 17173 from \citet{car02} 
(hereafter CHSM~17173).
The spectrograph was operated in the cross-dispersed mode with the
31.7~l~mm$^{-1}$ grating and a $0\farcs3\times6\arcsec$ slit, producing
full coverage from 1 to 2.5~\micron\ with a spectral resolution that varied
between $R=\lambda/\Delta\lambda=1800$ and 2200 depending on wavelength. 
Each object 
was dithered between two positions along the slit separated by $3\arcsec$.
The numbers of exposures and the integration times were $8\times60$, 
$8\times100$, and $20\times100$~sec for CHSM~17173, OTS~44, and OTS~7, 
respectively.
A nearby A0~V star (HD~98671) was also observed for the correction of telluric
absorption. After dark subtraction and flat-fielding, adjacent images
along the slit were subtracted from each other to remove sky emission.  
The sky-subtracted images were aligned and combined.  
A spectrum was extracted from each of the five orders in the cross-dispersed
format. The detectable hydrogen absorption lines in the spectrum of the 
telluric standard were removed manually through interpolation.
The intrinsic spectral slope of the standard was removed with an artificial 
blackbody spectrum of $T_{\rm eff}=10000$~K. At a given order, this modified 
spectrum was then divided into the extracted spectra of the three targets. 
Wavelength calibration was performed with OH airglow lines. 
For each object, the spectra from the five orders were scaled 
multiplicatively to align the overlapping regions between adjacent orders
to the same flux level.

During the spectral classification of the GNIRS targets in \S~\ref{sec:mem},
we use spectra of standards obtained with the near-IR 
spectrometer SpeX \citep{ray03} at the NASA Infrared Telescope Facility (IRTF).
The field dwarf LHS~2065 (M9V) and the young Taurus member KPNO~4 
(M9.5, \citet{bri02}) 
were observed with SpeX on the nights of 2003 December 21 and 23, respectively.
The instrument was operated in the prism mode with a $0\farcs5$ slit, 
producing a wavelength coverage of 0.8-2.5~\micron\ and a resolution of 
$R\sim200$. The spectra were reduced with the Spextool package \citep{cus04},
which included the same basic steps described for the GNIRS data reduction.
We corrected for telluric absorption with the method described by \citet{vac03}.

\section{Analysis}

\subsection{Membership and Spectral Classification}
\label{sec:mem}

\begin{figure}
\plotone{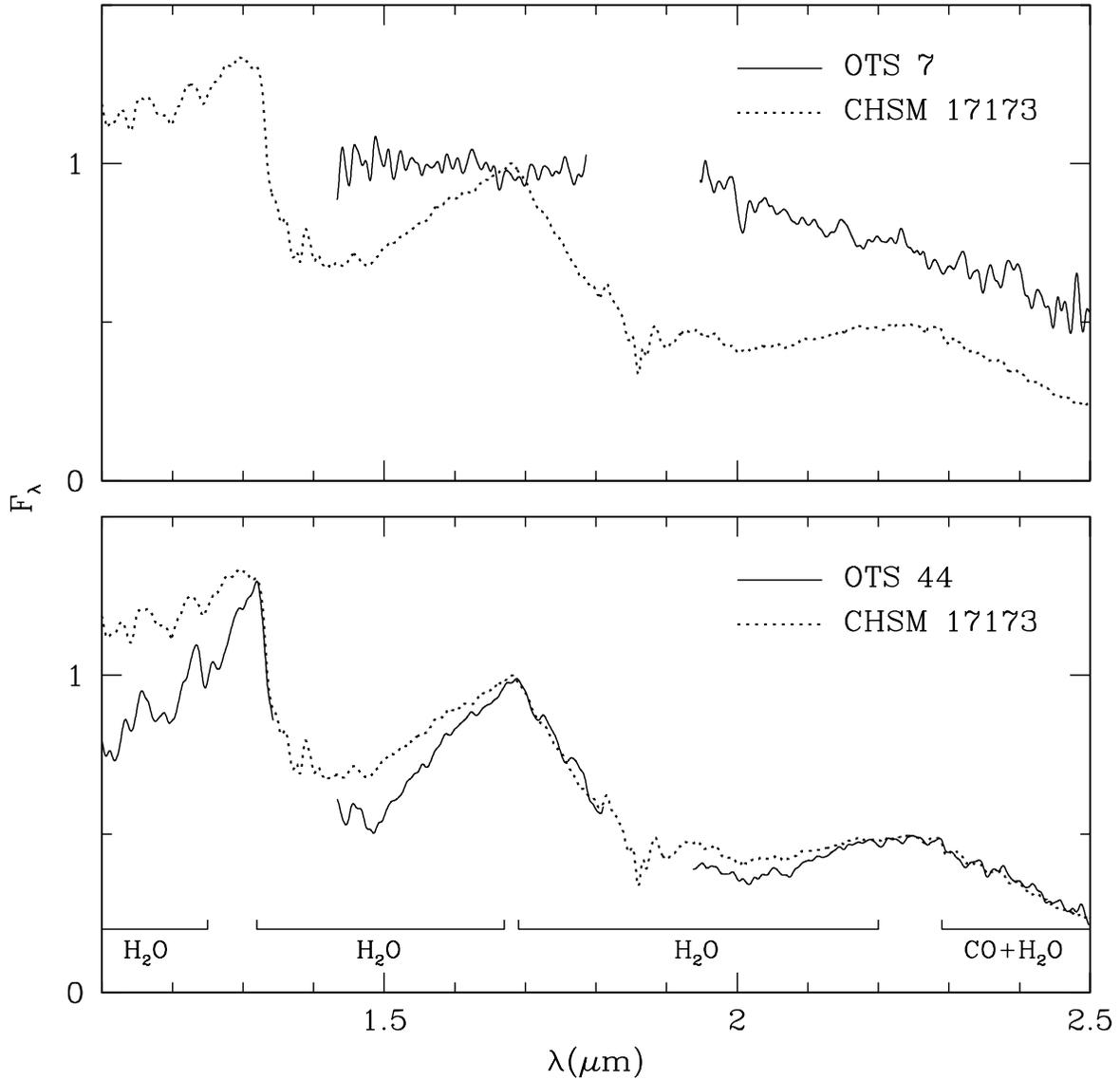}
\caption{
GNIRS near-IR spectra of two candidate substellar members of the Chamaeleon I
star-forming region identified by \citet{ots99} (OTS~7 and 44). The 
previously known late-type member CHSM 17173 (M8) is included for comparison.
The steam absorption bands in the spectrum of OTS~44 confirm that it is a
cool object and indicate a spectral type later than that of CHSM 17173.
The other candidate, OTS~7, is probably a field star or extragalactic source
given the absence of steam absorption or other late-type features.
The spectra are displayed at a resolution of $R=200$ and are 
normalized at 1.68~\micron.
}
\label{fig:spec1}
\end{figure}

The GNIRS spectra of the brown dwarf candidates OTS~7 and 44 and the known 
late-type Chamaeleon member CHSM~17173 are shown in Figure~\ref{fig:spec1}.
The spectrum of OTS~44 closely resembles that of CHSM~17173, mostly notably
in terms of the strong steam absorption bands. In this section,
we assess the membership of this object in Chamaeleon~I and measure its
spectral type. Meanwhile, the steam absorption expected from a cool photosphere
is absent from the spectrum of OTS~7. This object is probably 
an early-type field star or a background galaxy and is discussed no further.

\begin{figure}
\plotone{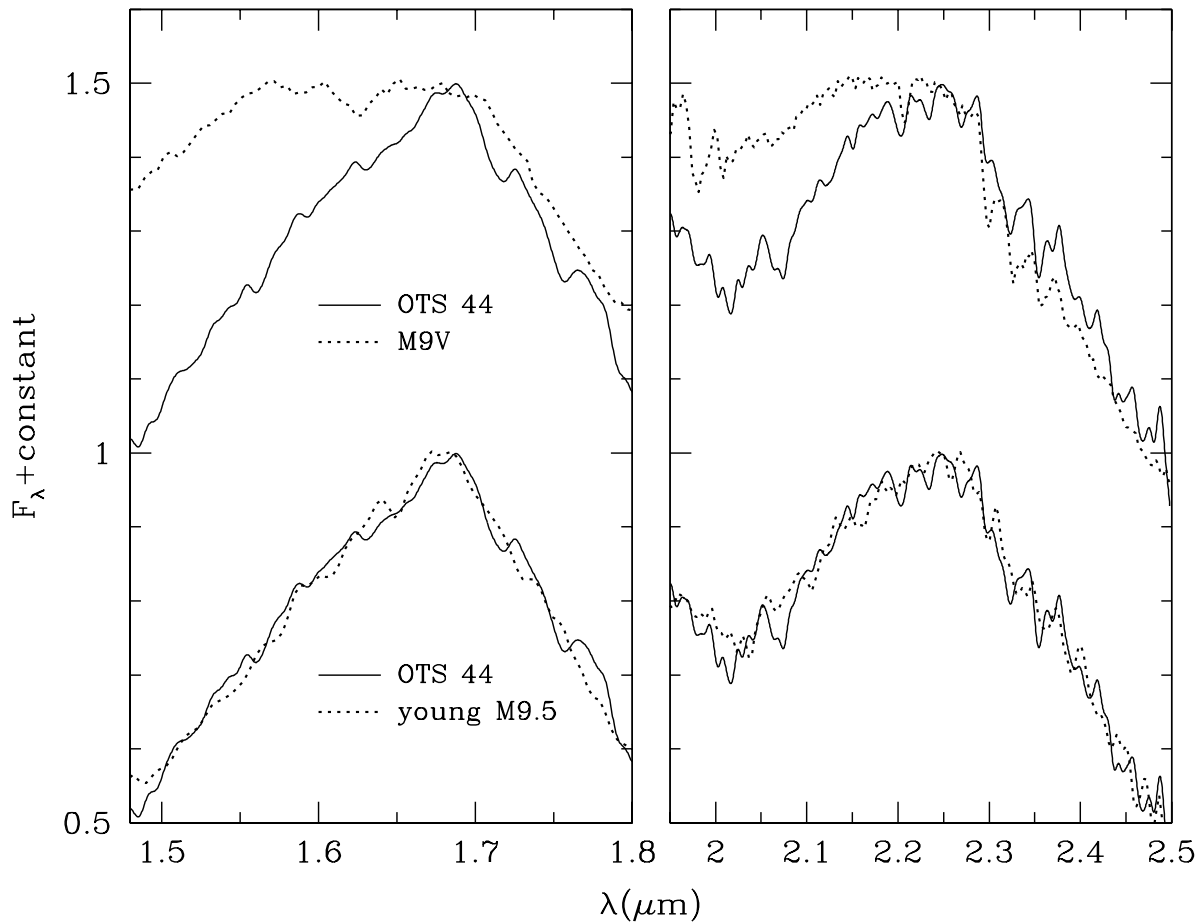}
\caption{
GNIRS $H$- and $K$-band spectra of the candidate brown dwarf OTS~44 
compared to IRTF SpeX data for the field dwarf LHS~2065 (M9V) and the young 
Taurus member KPNO~4 (M9.5). OTS~44 exhibits the triangular continua that are
indicative of young late-type objects rather than the broad plateaus 
found in field dwarfs \citep{luc01}, confirming its youth and membership 
in the Chamaeleon~I star-forming region. A spectral type of M9.5 is assigned 
to OTS~44 based on the close match to KPNO~4.
The spectra are displayed at a resolution of $R=200$ and are 
normalized at 1.68 and 2.25~\micron.}
\label{fig:spec2}
\end{figure}

OTS~44 could be a field dwarf, a field giant, or
a young, pre-main-sequence member of Chamaeleon~I. 
These distinct luminosity classes can be differentiated with spectral features 
that are sensitive to surface gravity, a variety of which have
been identified for cool objects at both optical 
\citep{mar96,luh99,mc04} and near-IR \citep{luh98,gor03,mc04} wavelengths.
In this work, we do not consider the gravity-sensitive atomic transitions, 
such as K~I at $J$ and Na~I at $K$, because of insufficient signal-to-noise
and spectral resolution in the GNIRS spectrum of OTS~44 and the SpeX data, 
respectively. Instead, we use the steam absorption bands to determine 
if OTS~44 is a young member of Chamaeleon~I. The shape of the continuum 
induced by steam absorption varies noticeably with surface gravity.
The broad plateaus in the $H$ and $K$ spectra of late-M and L dwarfs
\citep{rei01,leg01} are absent in young objects, 
resulting in sharply peaked, triangular continua \citep{luc01}.
This gravity diagnostic is easily detectable at the signal-to-noise and
resolution of our data, as illustrated with the spectra of the young late-type 
object KPNO~4 and the field dwarf LHS~2065 in Figure~\ref{fig:spec2}.
OTS~44 is compared to these two objects in Figure~\ref{fig:spec2}, where
we find that it clearly exhibits the triangular continua that are
indicative of youth rather than the broad plateaus expected for a field dwarf.
Meanwhile, the strength of the CO bandheads in OTS~44 is comparable to that
of the comparison dwarf and young object and much less than that of
an M giant (e.g., \citet{lr99}). 
Based on the behavior of these spectral features, we conclude that 
OTS~44 is not a field dwarf or giant and is instead a member of the
Chamaeleon~I star-forming region.

In addition to surface gravity, the IR steam bands are also sensitive to
temperature and thus can be used to measure spectral types
\citep{wgm99,cus00,rei01,leg01,tes01}.
At a given optical spectral type, these bands are stronger in young objects
than in field dwarfs \citep{lr99,luc01,mc04}, as shown with
KPNO~4 and LHS~2065 in Figure~\ref{fig:spec2}.  As a result, 
if a young source is classified by comparing the strength of its steam 
absorption to that of field dwarfs, the derived spectral type will be
systematically too late.
To arrive at accurate spectral types, optically-classified young objects
rather than dwarfs should be used as the standards \citep{lr99,luh03b}, 
which is the approach we adopt for classifying OTS~44. 
In Figure~\ref{fig:spec2}, we find that the depths of the steam bands in 
OTS~44 closely match those in KPNO~4, which has an optical type of M9.5. 
Because few IR spectra of young optically-classified late-type objects 
are available, we cannot compare OTS~44 to a finely sampled sequence of 
standards and determine the precise range of types that would match the steam
depths for OTS~44. This source is clearly later than CHSM~17173 
(M8, \citet{luh04a}) according to the relative
strengths of their steam bands, as shown in Figure~\ref{fig:spec1}.
Meanwhile, we have no constraints on the latest possible type for this object.
For the purposes of this work, we assign an uncertainty of $\pm1$~subclass
to the M9.5 classification of OTS~44.
The latest spectral types for previously confirmed members of Chamaeleon~I
are M8 for CHSM~17173 \citep{luh04a}, M8.25 for 2MASS~J11011926-7732383B 
\citep{luh04b}, and M8.5 for source 554 from 
\citet{com04}\footnote{
The spectral types from \citet{com04} are systematically later than 
those in \citet{luh04a} by an average of 0.5~subclass, which suggests
that 554 is probably M8 in our spectral type system.}. At a type of M9.5,
OTS~44 is now the latest confirmed member of Chamaeleon~I.

\subsection{Extinction, Temperature, and Luminosity}
\label{sec:ext}

We can estimate the extinction of OTS~44 from its near-IR spectrum and colors.
The comparison of the 1-2.5~\micron\ spectra of OTS~44 and CHSM~17173
in Figure~\ref{fig:spec1} and the relative $J-H$ and $H-K_s$ colors of 
OTS~44 and KPNO-Tau~4 imply that OTS~44 is slightly redder ($A_J\sim0.3$)
than both of these objects. Both CHSM~17173 and KPNO~4 lack noticeable
reddening in their optical spectra. Therefore, for OTS~44 we estimate an 
extinction $A_J=0.3\pm0.3$.
For the M9.5 spectral type of OTS~44, we adopt the 
temperature of 2300~K used by \citet{bri02} for KPNO~4. 
The bolometric luminosity is estimated by combining the $H$-band measurement 
from \citet{ots99}, a distance of 168~pc \citep{whi97,wic98,ber99}, and a
bolometric correction for M9.5 from \citet{rei01}.
The combined uncertainties in $A_H$, $H$, BC$_H$, and the distance modulus 
($\sigma\sim0.2$, 0.05, 0.2, 0.13) correspond to an error of $\pm0.12$ 
in log~$L_{\rm bol}$. The extinction, effective temperature, and bolometric 
luminosity for OTS~44 are listed in Table~\ref{tab:data}.
For comparison, we also include the same parameters for KPNO~4 as
estimated by \citet{bri02}.

\subsection{Mass}
\label{sec:mass}

\begin{figure}
\plotone{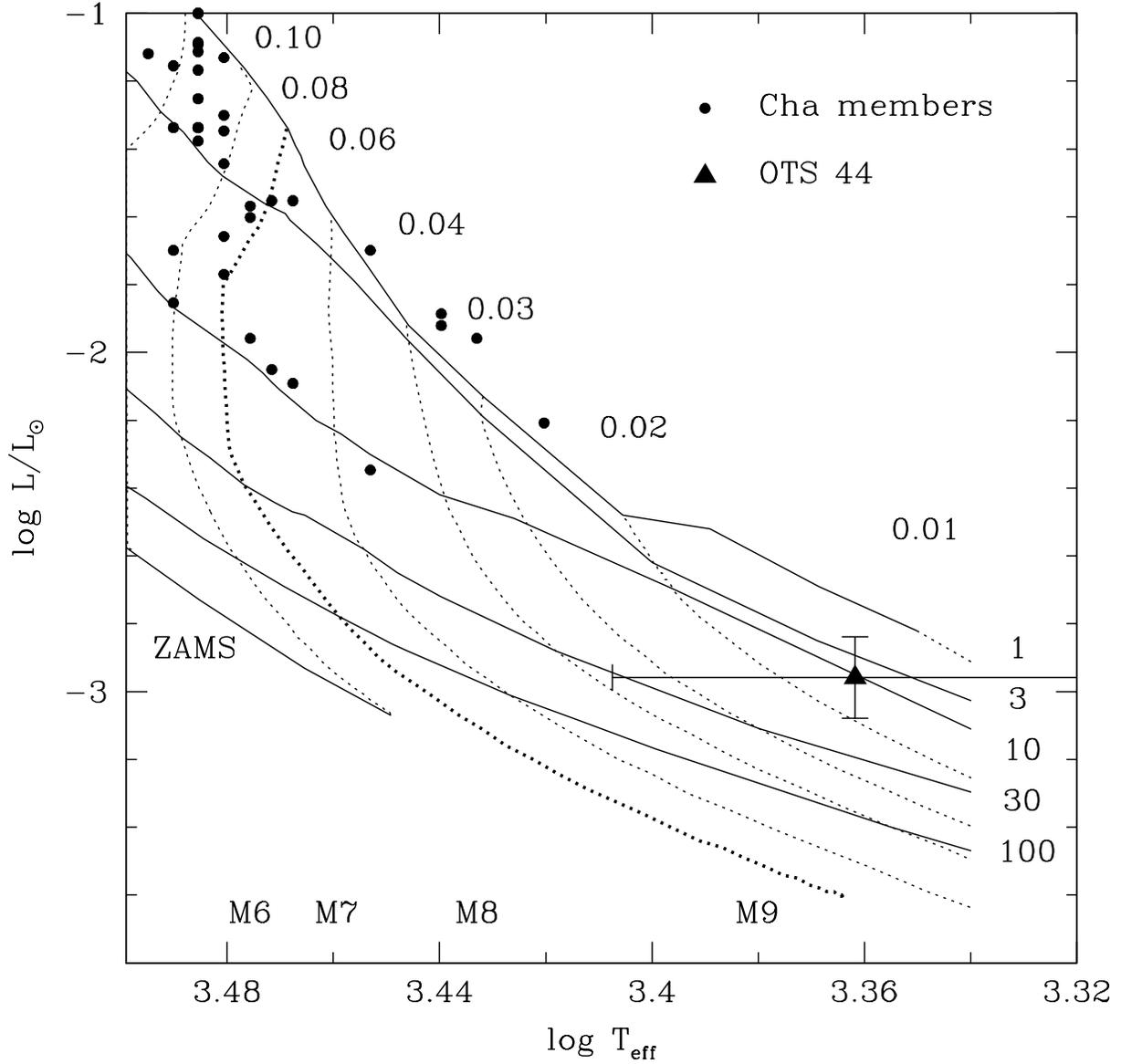}
\caption{
H-R diagram for previously known late-type members of Chamaeleon~I 
\citep{luh04a,luh04b} and the new member OTS~44
shown with the theoretical evolutionary models of
\citet{bar98} ($M/M_\odot>0.1$) and \citet{cha00} ($M/M_\odot\leq0.1$),
where the mass tracks ({\it dotted lines}) and isochrones ({\it solid lines}) 
are labeled in units of $M_\odot$ and Myr, respectively. 
These models imply a mass of $\sim0.015$~$M_\odot$ for OTS~44. 
}
\label{fig:hr}
\end{figure}

The temperature and luminosity for OTS~44 from the previous section 
can be used to estimate its mass via theoretical evolutionary models.
We select the models of \citet{bar98} and \citet{cha00} because they provide 
the best agreement with observational constraints \citep{luh03b}.
OTS~44 and previously known late-type members of Chamaeleon~I 
\citep{luh04a,luh04b} are plotted on the Hertzsprung-Russell (H-R) 
diagram in Figure~\ref{fig:hr} with these evolutionary models, which 
imply a mass of $\sim0.015$~$M_\odot$ for OTS~44.
The uncertainty in the spectral type of OTS~44 corresponds to roughly a 
factor of two in mass. The conversion of spectral type to temperature and
the models themselves contribute additional errors to this mass estimate 
that cannot be quantified. Nevertheless, because OTS~44 has the lowest
luminosity and temperature of any confirmed member of Chamaeleon~I, 
it very likely has the lowest mass. This brown dwarf is one of the nearest and
least massive free-floating objects discovered to date.

\acknowledgements
We thank the staff of Gemini Observatory, particularly Bernadette Rodgers,
for performing the GNIRS observations. We also thank John Rayner for assistance
with the IRTF SpeX observations. K. L. was supported by grant NAG5-11627 
from the NASA Long-Term Space Astrophysics program.

\clearpage
\begin{deluxetable}{llllllllll}
\tabletypesize{\scriptsize}
\tablecaption{Data for OTS 44 and KPNO-Tau 4\label{tab:data}}
\tablehead{
\colhead{} &
\colhead{} &
\colhead{} &
\colhead{} &
\colhead{$T_{\rm eff}$} &
\colhead{} &
\colhead{} &
\colhead{} &
\colhead{} &
\colhead{} \\
\colhead{ID} &
\colhead{$\alpha$(J2000)} & 
\colhead{$\delta$(J2000)} &
\colhead{Spectral Type} &
\colhead{(K)} &
\colhead{$A_J$} & 
\colhead{$L_{\rm bol}$} & 
\colhead{$J-H$} & 
\colhead{$H-K_s$} & 
\colhead{$K_s$}} 
\startdata
OTS 44 & 11 10 09.33\tablenotemark{a} & -76 32 18.1\tablenotemark{a} & M9.5$\pm1$ & 2300\tablenotemark{b}  & 0.3 & 0.0013 & 1.01\tablenotemark{c} & 0.79\tablenotemark{c} & 14.61\tablenotemark{c} \\
KPNO-Tau 4\tablenotemark{d}  &    04 27 28.01 &    26 12 05.3 &  M9.5$^{+0.5}_{-0.25}$ &   2300 &    0.00 &  0.0023 &     0.97      &  0.74      & 13.28      \\

\enddata
\tablenotetext{a}{Measured in $I$-band images from Luhman (in preparation).}
\tablenotetext{b}{Temperature for M9.5 from \citet{bri02}.}
\tablenotetext{c}{\citet{ots99}.}
\tablenotetext{d}{Data for KPNO-Tau 4 from \citet{bri02}, updated with the 
values of $JHK_s$ from the 2MASS Point Source Catalog.}
\end{deluxetable}

\clearpage

\end{document}